\documentclass[twocolumn]{aastex6}

\usepackage{graphicx}
\usepackage[flushleft]{threeparttable}
\usepackage{blindtext}
\usepackage{amsmath}
\usepackage{mathtools}
\usepackage{multirow}
\usepackage{hyperref}
\hypersetup{
	colorlinks	= true,
	linkcolor	= red,
	urlcolor	= cyan,
	citecolor	= blue
}

\begin{document}

\title{Exploring Biases of Atmospheric Retrievals in Simulated \emph{JWST} Transmission Spectra of Hot Jupiters}

\author{M. Rocchetto$^1$ , I. P. Waldmann$^1$, O. Venot$^2$, P.-O. Lagage$^{3,4}$, G. Tinetti$^1$}
\affil{$^1$ Department of Physics \& Astronomy, University College London, Gower Street, WC1E6BT London, United Kingdom}
\affil{$^2$ Instituut voor Sterrenkunde, Katholieke Universiteit Leuven, Celestijnenlaan 200D, 3001 Leuven, Belgium}
\affil{$^3$ Irfu, CEA, Universit\'{e} Paris-Saclay, F-9119 Gif-sur Yvette, France}
\affil{$^4$ AIM, Universit\'{e} Paris Diderot, F-91191 Gif-sur-Yvette, France}

\email{m.rocchetto@ucl.ac.uk}

\begin{abstract}
With a scheduled launch in October 2018, the \emph{James Webb Space Telescope} (\emph{JWST}) is expected to revolutionise the field of atmospheric characterization of exoplanets.  The broad wavelength coverage and high sensitivity of its instruments will allow us to extract far more information from exoplanet spectra than what has been possible with current observations. In this paper, we investigate whether current retrieval methods will still be valid in the era of \emph{JWST},  exploring common approximations used when retrieving transmission spectra of hot Jupiters. 
To assess biases, we use 1D photochemical models to simulate typical hot Jupiter cloud-free atmospheres and generate synthetic observations for a range of carbon-to-oxygen ratios.  Then, we retrieve these spectra using TauREx, a Bayesian retrieval tool, using two methodologies: one assuming an isothermal atmosphere, and one assuming a parametrized temperature profile. Both methods assume constant-with-altitude abundances. 
We found that the isothermal approximation biases the retrieved parameters considerably, overestimating the abundances by about one order of magnitude. The retrieved abundances using the parametrized profile are usually within one sigma of the true state, and we found the retrieved uncertainties to be generally larger compared to the isothermal approximation. Interestingly, we found that using the parametrized temperature profile we could place tight constraints on the temperature structure. This opens the possibility to characterize the temperature profile of the terminator region of hot Jupiters.
Lastly, we found that assuming a constant-with-altitude mixing ratio profile is a good approximation for most of the atmospheres under study.

\end{abstract}

\maketitle
 
\section{Introduction} 
\label{sec:introduction}

The discovery of over three thousands exoplanets has unveiled a large and diverse population. We see planets in a range of sizes, temperatures, and orbits, far exceeding the diversity found in our own Solar System. Today, emphasis in the field of exoplanets is shifting from the \emph{discovery} to the \emph{characterization} of these exoplanetary bodies, as understanding their nature will in turn provide important clues on the planet's formation and evolution history. 

The study of exoplanetary atmospheres represents one of the most immediate and direct ways to characterize exoplanets. To date, the atmospheres of several tens of giant planets, sub-Neptunes and super-Earths have been studied and characterized with the  \emph{Hubble Space Telescope} (\emph{HST}) \citep{Charbonneau:2002er, Tinetti:2010cl,Swain:2013hn,Kreidberg:2014fn,Sing:2016hi,Tsiaras:2016kh} \emph{Spitzer} space telescope  \citep{Tinetti:2007hy,Allen:2007bl,Grillmair:2007ee,Charbonneau:2008ep,Beaulieu:2010gl,Stevenson:2010hf,Knutson:2011bj,Deming:2011fs,Todorov:2013be} and other ground based facilities  \citep{Redfield:2008jo,Snellen:2008dq,Griffith:2010gu,Waldmann:2012jx,Bean:2013dg,Waldmann:2014fc,Zellem:2014ey,Brogi:2014kz}. 

With the imminent launch of the \emph{JWST}, it has become fundamental to assess whether the current methods used to interpret these spectra will still be valid when higher quality datasets will be available. In this work we aim to answer this question in part,  exploring  the biases induced by common assumptions used in atmospheric retrievals.

One of the major limitations of current observations is the limited wavelength coverage.  The best quality datasets, which led to the confident  detection of water vapor in several hot Jupiters and warm Neptunes, have been mainly obtained with the Wide Field Camera~3 onboard \emph{HST}, covering the spectral range 1.1--1.7 $\mu$m.  Nevertheless, it is at longer wavelengths that most roto-vibrational transitions of molecular species occur. While the \emph{Spitzer Space Telescope} has  given some insight into the longer wavelength regime to several tens of close-in hot Jupiters, the data have relatively large uncertainties, and they are mostly only photometric measurements. Significant  advances in the field of atmospheric characterization can therefore only happen if high quality observations extending to the longer wavelength regime are obtained.

In this scenario, the \emph{James Webb Space Telescope} (\emph{JWST}) will undoubtedly revolutionise the field of exoplanetary atmospheres, addressing two major problems affecting current observations: wavelength coverage and instrument sensitivity. With a scheduled launch for 2018 October, the large spectral coverage (0.7--28 $\mu$m) covered by its multiple instruments, combined with high sensitivity and high degree of instrumental characterization and calibration, will ensure a significant advance in atmospheric characterization  \citep{Beichman:2014gj,Cowan:2015bw,Batalha:2015vb,Greene:2016gx,Barstow:2015br,Barstow:2016ti}.

Atmospheric spectra of transiting exoplanets in a broad spectral range will enable us to constrain the abundances of different molecular species, the temperature structure of the atmosphere, and the presence or absence of clouds and hazes. In the case of warm H/He dominated atmospheres one of the key elemental ratio that we aim to constrain is the carbon-to-oxygen ratio (C/O). Such measurements will enable us to distinguish between different  formation and migration scenarios, so far  poorly constrained  \citep{Oberg:2011je,Madhusudhan:2011kw,AliDib:2014ki,Thiabaud:2015dk}. While transmission and emission spectra do not provide direct constraints on the elemental abundances, the measurement of the absolute abundances of O-bearing and C-bearing molecules will provide some constraints on the C/O ratio. In particular, the excess carbon and oxygen not locked in CO will form  either oxygen-bearing molecules such as H$_2$O in atmospheres with  C/O $<$ 1, or, in atmospheres with C/O $>$ 1, carbon-rich species such as HCN, C$_2$H$_2$  and CH$_4$ \citep{Madhusudhan:2012ga,Moses:2013jh,Fortney:2013jc,Venot:2015gi}.  Determining the atmospheric abundances of these gases in hot Jupiters with high accuracy is therefore paramount and   \emph{JWST} will give us direct access to absorption features of these molecules both in emission and transmission. 

Determining the absolute abundances of atmospheric gases from atmospheric spectra requires the use of retrieval methods. Atmospheric retrieval techniques are now commonly used to infer the properties of exoplanetary atmospheres, including molecular abundances and temperature profiles  \citep{Madhusudhan:2009gd,Madhusudhan:2011ex,Benneke:2012ig, Benneke:2013hu,Lee:2011gl,Line:2012ki,Line:2013ej,Line:2014eu,Irwin:2008dp, deWit:2013kz,Waldmann:2015ev,Waldmann:2015iq}.  These tools enable us to fully map the likelihood space of atmospheric models, and  to place upper limits and constraints on the abundances of molecules and temperature profiles.

The lack of high signal-to-noise and broad wavelength coverage observations have however led current retrievals and forward models to make several assumptions and  approximations to reduce the parameter space. The forward model included in most retrieval methods is a 1D radiative transfer model  \citep{Brown:2001ck,Seager:2011dl,Tinetti:2012jd,Liou:2002uh,Hollis:2013ea}, implementing opacity cross sections  for the major molecular absorbers, Rayleigh scattering and collision induced absorption. Transmission spectra are usually retrieved assuming constant-with-altitude temperature and molecular abundances. This might be a fair approximation when probing narrow wavelength ranges, but can lead to significant biases when larger wavelength ranges are probed. \emph{One of the pressing questions we are facing today is whether these assumptions will still be valid in the era of \emph{JWST}}.

In this paper, we aim to address these issues. We study the biases and degeneracies of atmospheric retrievals of high quality, broad wavelength range transmission spectra of hot Jupiters, such as those that will be obtained with instruments onboard \emph{JWST}. We apply and compare different retrieval approaches to synthetic  observations for a range of hot atmospheres with different C/O  computed using photochemical models, and study the biases of common assumptions used in today's retrievals. 

This study aims at answering the following questions:
\begin{itemize}
\item[a)] Are our retrieval approaches and forward models appropriate for the high signal-to-noise, and broader wavelength range spectra expected from future facilities such as  \emph{JWST}?
\item[b)] Can we confidently retrieve absolute molecular abundances and infer the C/O ratio?
\end{itemize}

In Section 2 we describe the chemical and radiative transfer models used to generate the synthetic transmission spectra. We also present the \emph{JWST} synthetic observations, and describe the two retrieval approaches used to interpret these synthetic observations. In Section 3 we describe qualitatively the simulated transmission spectra and present the results of the retrievals. In Section 4 we discuss our results, and in Section 5 we summarize the main conclusions of this study.

\section{Method}

\subsection{Chemical models}
The 1D atmospheric chemical models were generated using the photochemical model developed for hot atmospheres \citep[][and references therein]{Venot:2012fr}. These models have been used to study exoplanets \citep{Venot:2014dk,Agundez:2014iu,Venot:2015gi,Venot:2015fg,Tsiaras:2016kh} as well as Solar System giant planets \citep{Cavalie:2014et,Mousis:2014ii}. The chemical scheme has been developed with combustion specialists and  validated in a wide range of pressures (0.001--100 bar) and temperatures (300--2500 K), making this model one of the currently most reliable chemical schemes \citep{BattinLeclerc:2006tv, Bounaceur:2007kv, Anderlohr:2010fn, Wang:2010en}. \cite{Venot:2015gi} showed that the use of more complete chemical models, including species with up to six carbon atoms, has little effect on the synthetic spectra. We therefore used the simpler, and computationally faster, scheme which includes species with up to four carbon atoms and is able to model the kinetic behaviour of species with up to two carbon atoms.  This scheme includes 105 neutral species and 960 reactions (and their reverse reactions). We used a constant diffusion coefficient, $K_\mathrm{zz} = 10^8$ cm$^2$s$^{-1}$ due to the uncertainties on the vertical mixing acting in exoplanet atmospheres. A similar value has been often used in the literature  \citep{Lewis:2010bt,Moses:2011bn,Line:2011ih,Venot:2013kv}. We note however that this value might be too high, as pointed out by \cite{Parmentier:2013fd}.

\begin{figure}
	\centering
	\includegraphics[width=150pt]{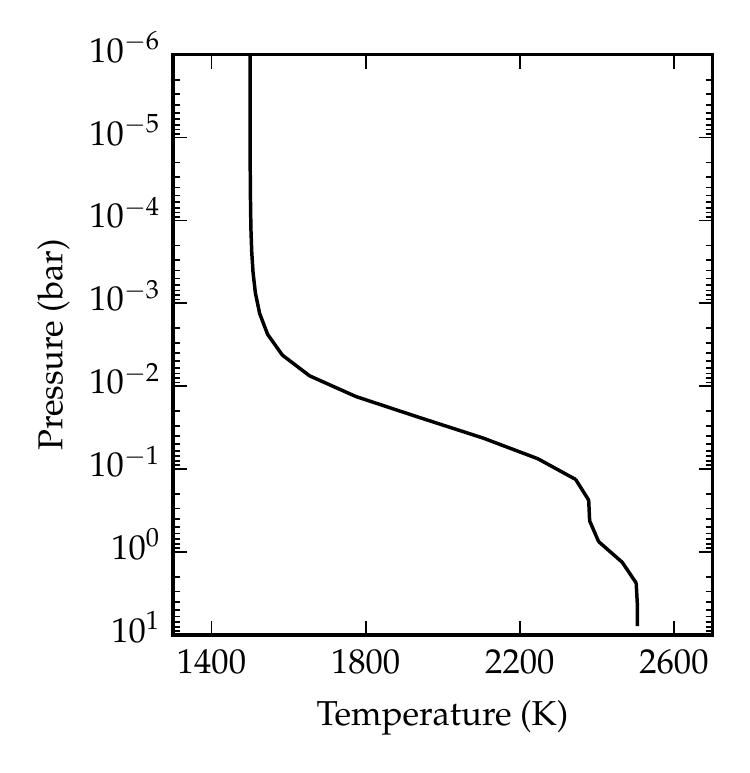}
	\caption{Temperature-pressure profile used for the atmospheres under study.}
	\label{fig:tph_profile}
\end{figure}

We used a temperature-pressure (TP) profile with a high-altitude temperature of  1500 K. The vertical profile is the same as the one used in \cite{Venot:2015gi}. It was computed using the analytical model one of \cite{Parmentier:2014ju}, using coefficients from \cite{Parmentier:2015ky} and the opacities from \cite{Valencia:2013ce}. The profile, shown in Figure \ref{fig:tph_profile}, was obtained by setting the irradiation temperature to 2300 K and the internal temperature $T_\mathrm{int}  = 100\,$K. We assumed a planet with $R_p = 1.162\,_J$ and $M_p = 1.138\,M_J$. 

We computed chemical models for an atmosphere of solar metallicity with  C/O of 0.5, 0.7, 0.9, 1.0, 1.1, 1.3, and 1.5. 

\subsection{Synthetic high resolution transmission spectra}

High resolution (R $\approx$ 10000) synthetic transmission spectra were computed using the  forward models included in TauREx \cite{Waldmann:2015iq}. This forward model is based on a 1D radiative transfer model that calculates the optical path through the planetary atmosphere. It results in a transmission spectrum of transit depth, i.e. $(R_p/R_*)^2$, as a function of wavelength. The temperature profile used is the same as the one used for the computation of the photochemical models (Figure \ref{fig:tph_profile}). We include a precise computation of the pressure-altitude profile, and take into account the effect of gravity, temperature and mean molecular weight in the computation of the scale height in each of the 100 atmospheric layers included in the model. We compute the pressure grid from $10^{-4}$ to 10 bar, and define the 10 bar pressure radius to be $R_p = 1.162\,R_J$. The mass is set to  $M_p = 1.138\,M_J$. Amongst the 105 molecules considered in the photochemical model we only consider the following seven molecules in the computation of the opacity in the synthetic spectra: C$_2$H$_2$, CH$_4$, CO, CO$_2$, H$_2$O, HCN and NH$_3$. We found that amongst the complete set of 105 molecules contained in the chemical model, these are the most abundant ones in all cases and will therefore dominate the spectral modulation.  The wavelength dependent cross sections for these absorbing molecules were computed using line lists from ExoMol \citep{Barber:2006tl,Harris:2006eb,Yurchenko:2011hs,Tennyson:2012ca,Yurchenko:2013jd,Barber:2014jm}, HITRAN \citep{Rothman:2013em} and HITEMP \citep{Rothman:2010in}. Note that the mean molecular weight of each atmospheric layer is coupled to the mixing ratio of all 105 molecules. We included additional opacity from Rayleigh scattering of H$_2$ and from collision induced absorption of  He and H$_2$-H$_2$ and H$_2$-He  pairs \citep{Richard:2012jl}. 

\subsection{\emph{JWST} spectra}

\begin{table}
	\small
	\center
	\caption{\emph{JWST} instrument modes}
	\label{tab:jwst}
	\begin{tabular}{l l l  }
		\hline \hline
 		Instrument   & Mode    	&    Wavelength range ($\mu$m)	\\
		\hline		
		NIRISS 	& SOSS/GR700XD &  1.0--2.5 $\mu$m	\\
		NIRCam 	& LW grism/F322W2 &  2.5--3.9 $\mu$m	\\
		NIRCam 	& LW grism/F444W &  3.9--5.0 $\mu$m	\\
		MIRI 		& slitless/LRS prism &  5.0--10.0 $\mu$m	\\
\hline
       \end{tabular}
\end{table}

\begin{figure}
	\centering
	\includegraphics[width=\columnwidth]{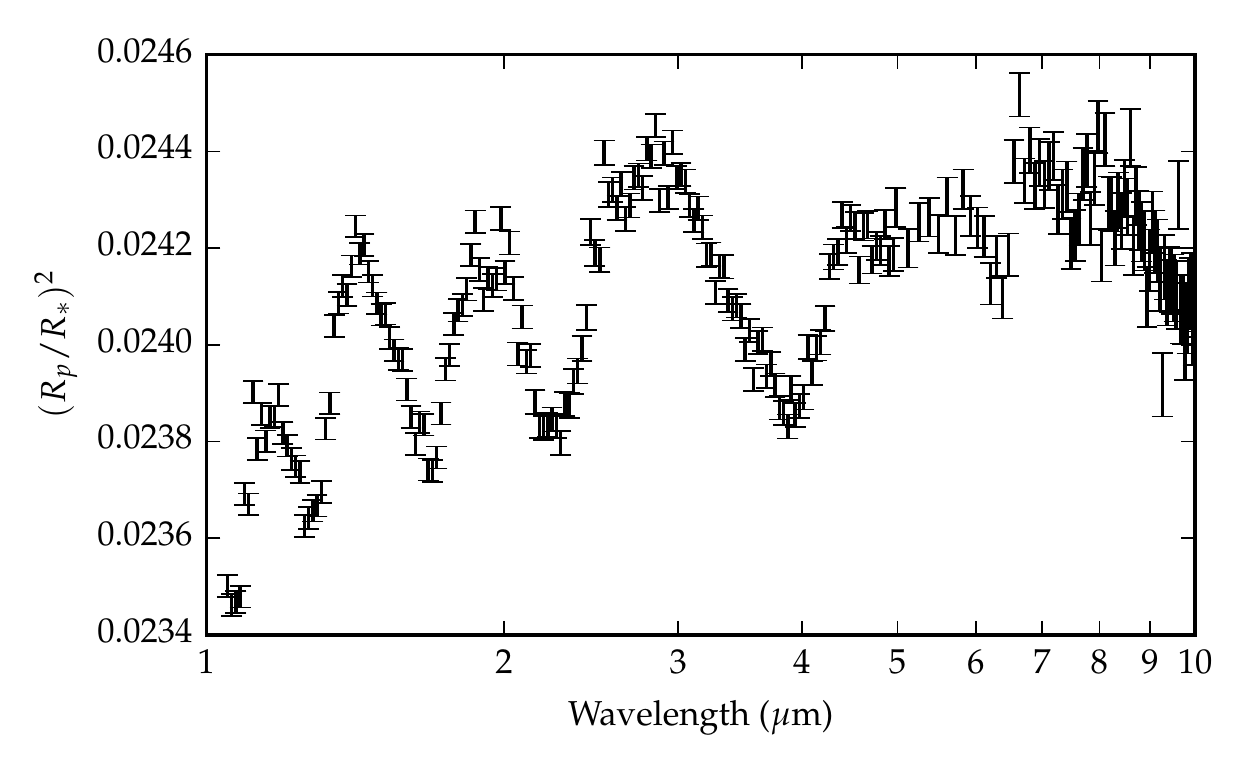}
	\caption{Simulated \emph{JWST} observation for C/O = 0.5. The spectrum was obtained combining four separate synthetic observations obtained with NIRISS, NIRCam and MIRI to cover the 1--10  $\mu$m spectral range. This spectrum would therefore require observing a total of four transits.}
	\label{fig:jwst_spectrum}
\end{figure}

We simulated spectra for the Near-InfraRed Imager and Slitless Spectrograph (NIRISS) in Single-Object Slitless Spectroscopy (SOSS) mode using the GR700XD optics \citep{Doyon:2012kn}. We applied a lower wavelength cutoff at 1 $\mu$m to avoid saturation and a long wavelength cutoff at 2.5 $\mu$m to avoid spectral contamination \citep{Greene:2016gx}. We then used the Near Infrared Camera (NIRCam) using the long wavelength (LW) channel and the F322W2 and F444W filters, covering the 2.5--3.9 and 3.9--5.0 $\mu$m spectral ranges respectively \citep{Greene:2007da}.  An alternative could be the use of the Near Infra Red SPECtrometer (NIRSPEC)  in its high resolution mode with the 2 instrumental configurations: F170LP/G235H (1.7--3.1  $\mu$m] and F290LP/G395H (2.9--5.2 $\mu$m) \citep{Ferruit:2014cs}.  Finally, we use the Mid Infrared Instrument (MIRI) to cover the 5.0--10.0 $\mu$m wavelength range. We use MIRI in slitless mode, using the Low Resolution Spectrometer (LRS) and we apply a long-wavelength cutoff of 10 $\mu$m due to the degrading S/N at longer wavelengths  \citep{Kendrew:2015jx}. Each observation covering the full wavelength range 1--10 $\mu$m will therefore require four separate observations. We have considered a one hour effective integration time during the transit and the same amount of time on the star alone. For each mode, the same amount of time was used. Table \ref{fig:jwst_spectrum} summarizes the instrument modes considered in this study. 

The noise in the spectra was calculated taking into account the star photon noise, the zodiacal and telescope background noise (integrated over the entire band pass of the spectrometer for the slitless mode), the detector dark current and noise. We assumed a star similar to HD189733. The star spectrum used was generated using the PHOENIX atmosphere star code \citep{Husser:2013ca}. For NIRISS and NIRCAM, we have binned the spectra to a constant spectral resolution of R = 100. For such a bright star we realized that we are in fact  very close to the limitation from systematics of the JWST. Such systematics are difficult to assess but we can reasonably assume that they will be lower than \emph{HST}. Given the latest performances achieved with HST  \citep[e.g.][]{Tsiaras:2016kh}, we can anticipate that the systematics for NIRISS and NIRCAM will be better than about 20 ppm. For MIRI, \cite{Greene:2016gx} adopted a value of 50 ppm, and \cite{Beichman:2014gj} took a value of 30 ppm; we have adopted an intermediate value of 40 ppm. An example of a final spectrum is shown in Figure \ref{fig:jwst_spectrum}. 

\subsection{Atmospheric Retrieval}

\begin{table*}
	\small
	\center
	\caption{Free parameters of the two retrieval approaches used in this study. The \textsc{tp-iso} approach refers to the retrieval using an isothermal TP profile, while the \textsc{tp-param} refers to the retrieval using a parametrized TP profile.}
	\label{tab:retrievals}
	\begin{tabular}{l l l l }
		\hline \hline
 		Approach   & Parameter    	&    Prior & Description	\\
		\hline
		
 		\textsc{tp-iso} & $\log$ H$_2$O, $\log$ CO, $\log$ CO$_2$,   &  $-12 \dots 1$    & Molecular abundances \\
 		 (10 free parameters)    & $\log$ CH$_4$ $\log$ NH$_3$,  $\log$ HCN,  &   &  \\
 		                                    & $\log$ C$_2$H$_2$ &            & \\
 		                                    & $T_\mathrm{iso}$ [K]									& $1300\dots2600$                      & Isothermal temperature \\
				                      & $R_p$ [$R_\mathrm{Jup}$]				              & 1.05\dots1.28   & Planetary radius at 10 bar \\
				                      & $\log(P_\mathrm{top}\,\mathrm{[Pa]})$								 & $0\dots6$  & Cloud top pressure \\
		\hline
 		\textsc{tp-param} & $\log$ H$_2$O, $\log$ CO, $\log$ CO$_2$,   &  $-12 \dots 1$    & Molecular abundances \\
 		 (14 free parameters)    & $\log$ CH$_4$ $\log$ NH$_3$,  $\log$ HCN,  &   &  \\
 		                                    & $\log$ C$_2$H$_2$ &            & \\
 		                                    & $T_\mathrm{irr}$ [K]						& $1300\dots2600$                        & Stellar flux at the top of the atmosphere \\
				                      & $\log \kappa_\mathrm{IR}$                                 & $-4\dots1$ 						& Mean infrared opacity 	\\			                 
				                      & $\log \kappa_{\nu1}, \log \kappa_{\nu2}$                & $-4\dots1$ 						& Optical opacity sources 	\\
				                      & $\alpha$								               & $0\dots1$ 					        	& Weighting factor for $\kappa_{\nu1}$and $\kappa_{\nu2}$ \\ 
				                      & $R_p$ [$R_\mathrm{Jup}$]				              & 1.05\dots1.28   & Planetary radius at 10 bar \\
				                      & $\log(P_\mathrm{top}\,\mathrm{[Pa]})$								 & $0\dots6$  & Cloud top pressure \\

\hline
       \end{tabular}
\end{table*}

The analysis and interpretation of the simulated observed spectra was carried out using TauREx \citep{Waldmann:2015iq}. Recently, TauREx  has been used to model the spectra obtained with the Wide Field Camera 3 onboard the Hubble Space Telescope for HD209458b and 55 Cnc e \citep{Tsiaras:2015tq,Tsiaras:2016kh}

Two retrieval approaches were used as part of the current study. Both approaches did not assume any prior knowledge on the chemistry, i.e. the absolute abundance of all gases taken into account is fitted independently. The only difference between the two approaches is in the parametrization of the temperature profile:

\begin{itemize}

\item In the first case we assumed an isothermal TP profile. We will refer to this method as ``\textsc{tp-iso}''. This approach is the most commonly used when fitting transmission spectra  \citep{Line:2012ki,Benneke:2012ig,Irwin:2008dp}, and includes a parametrization of the atmosphere assuming constant-with-altitude mixing ratio and temperature profiles. Crucially, it does not assume any prior on the chemistry of the atmosphere. The free parameters of the retrieval were the absolute abundance of each atmospheric constituent taken into account, the isothermal temperature, the cloud parameters and the 10 bar pressure radius. The mean molecular weight is coupled to the fitted composition, and we assumed the bulk atmosphere to be formed by a mixture of hydrogen and helium, whose ratio is fixed to solar value (85\% H$_2$ and 15\% He). We assumed uniform priors in log space for the absolute abundances, ranging from $10^{-12}$ to 1. We assumed uniform priors for the temperature (1300--2500) K and for the 10 bar radius (1.05--1.28 $R_\mathrm{Jup}$). The prior width of the 10 bar radius was determined by assuming a relative uncertainty on $R_p$ of 20\% ($R_p = 1.162$ $R_J$). Lastly, we fitted  the cloud top pressure with a uniform prior in log space (10$^{-5}$--10 bar). This parametrization resulted in 10 free variables. 

\item In the second case, we assumed a more complex TP profile described by five separate parameters. We will refer to this method as ``\textsc{tp-param}''. Since the temperature profile of the atmospheres under study is highly non-isothermal for pressures greater than 1 mbar (see Figure \ref{fig:tph_profile}), fitting an isothermal profile might lead to biases. We therefore investigated the effectiveness of fitting a more complex profile using this second method. We used the parametrization of \cite{Guillot:2010dd} modified by \cite{Line:2013ej} and \cite{Parmentier:2014ju}. There are five parameters that define the temperature profile: the planet internal heat flux ($T_\mathrm{int}$), the stellar irradiation flux ($T_\mathrm{irr}$), the opacities in the optical and infrared ($\kappa_{\nu1}, \kappa_{\nu2}$), and a weighting factor between optical opacities ($\alpha$).  For a full description of this model we refer the reader to Section 3.1 in \cite{Line:2013ej}.  These five parameters replace the single parameter used for the isothermal profile in the first method. This model only differs from the first one for the type of TP profile used. This parametrization resulted in 14 free variables.

\end{itemize}

The parametrized profile described above is commonly used in the retrieval of emission spectra, where the spectral features are more sensitive to temperature gradients than in transmission. It has received little attention in the retrieval of transmission spectra, as it is assumed that transmission spectra are much less sensitive to temperature gradients, and therefore isothermal profiles, thought to represent the ``average'' atmospheric temperature, have always been used. Previous studies have addressed the potential bias of the isothermal assumption \citep{Barstow:2013jc}, and found that some information on the temperature profile could be retrieved in transmission only in the highest signal-to-noise and broad wavelength coverage cases.

We used these two approaches to interpret the synthetic \emph{JWST} observations in a range of C/O. In all cases we used the MultiNest sampling algorithm  \citep{Feroz:2008fi} to finely sample the parameter space and obtain the posterior distributions of the model parameters. We chose this method instead of a more classical MCMC, as MultiNest can better map the likelihood of highly degenerate parameter spaces. Table \ref{tab:retrievals} summarises the free parameters and the corresponding prior widths used in the two  retrieval methods.

\section{Results}

\begin{figure*}
	\centering
	\includegraphics[width=500pt]{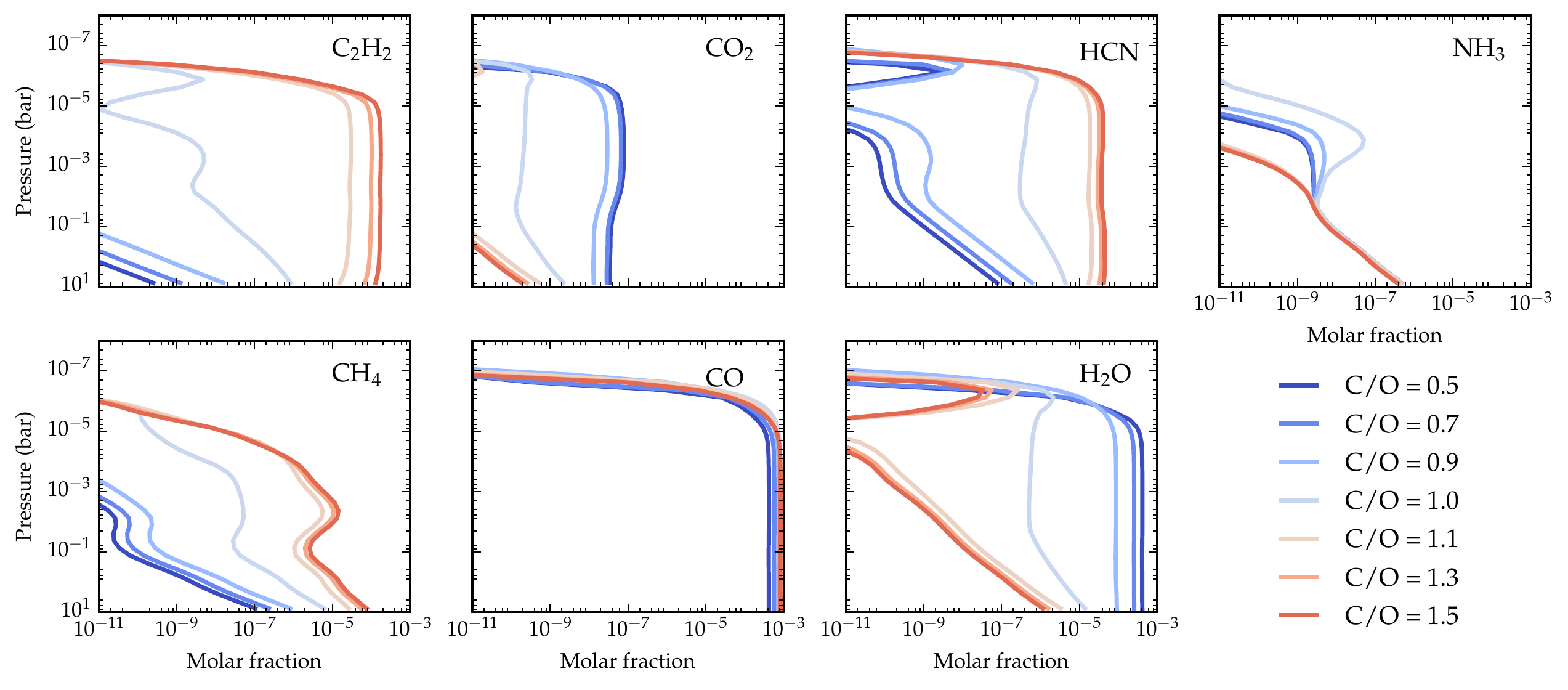}
	\caption{
	 Vertical abundance profiles for different molecules for a range of C/O. The different coloured lines show the molar fraction profiles at different C/O, as shown by the legend.}
	\label{fig:chemical_models}
\end{figure*}

\begin{figure*} 
	\centering
	\includegraphics[width=520pt]{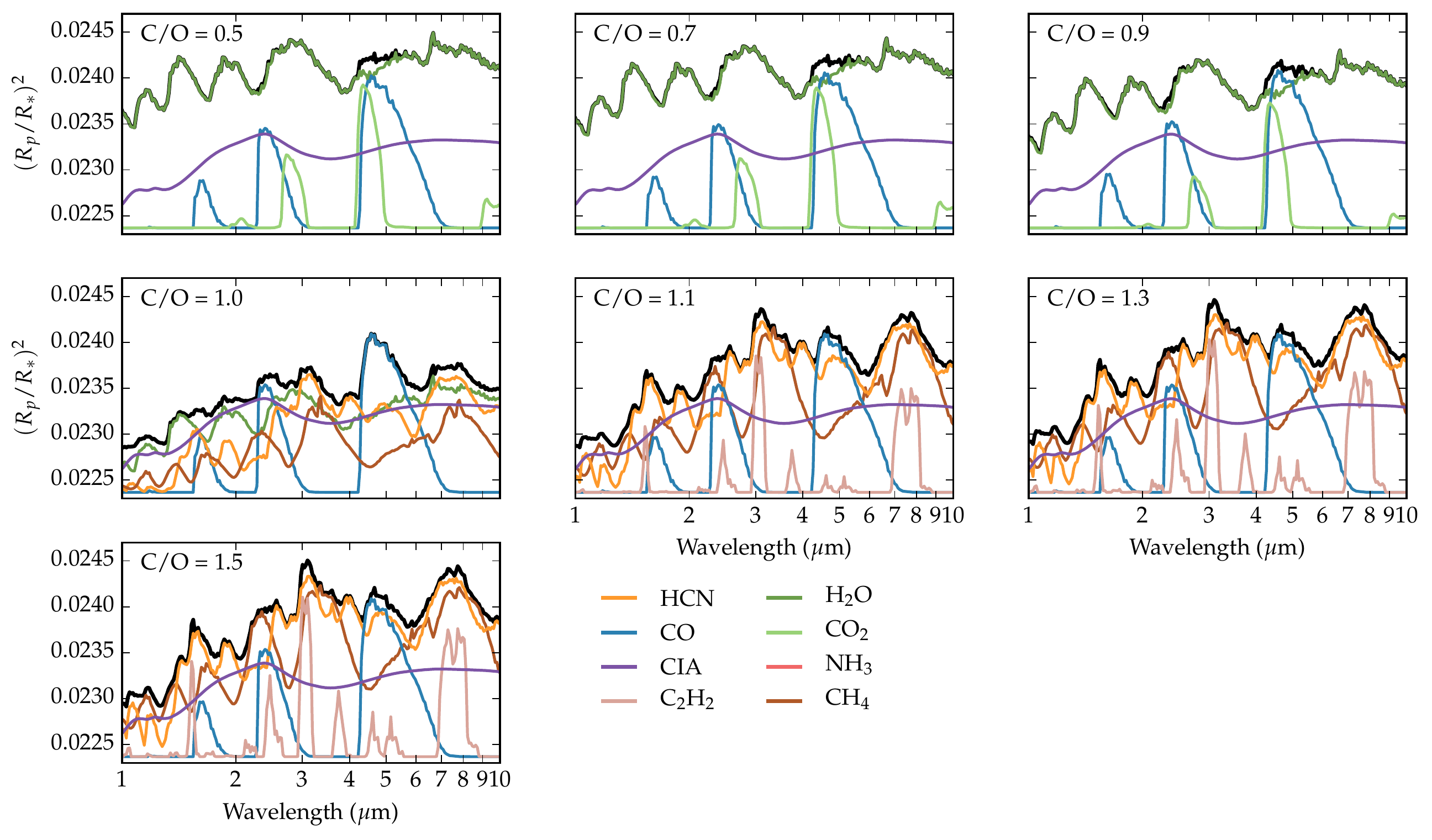}
	\caption{
	Synthetic transmission spectra (black lines) and contributions of the major opacity sources (colored lines, see legend) for the atmospheres whose chemistry is shown in Figure \ref{fig:chemical_models}, for different C/O values. The opacity sources include the seven molecules considered in this study, and the collision induced absorption (CIA) from H$_2$--H$_2$ and H$_2$--He pairs. Note that for each plot we only show the major opacity contributors to the spectrum, and we hide the molecules that do not significantly contribute to the transmission spectrum features.}
	\label{fig:spectra}
\end{figure*}

\subsection{Chemical models and transmission spectra} 

Figure \ref{fig:chemical_models} shows the vertical abundance profiles of seven molecules for all C/O ratios considered in this study, and Figure \ref{fig:spectra} shows the synthetic transmission spectra and contributions of the major opacity sources for the same C/O values. It can be clearly seen that the chemistry and the resulting spectra change significantly between C/O $<$ 1, C/O = 1 and C/O $>$ 1. 

Firstly, we note that while the transmission spectra of an oxygen rich atmosphere are dominated almost entirely by H$_2$O, with additional features from CO at 4.6  $\mu$m and from CO$_2$ at 4.3 $\mu$m, a carbon rich atmosphere is dominated by HCN and CH$_4$, with additional features from CO at 4.6 $\mu$m and C$_2$H$_2$  at 1.7, 3.0 and 7.5 $\mu$m. At the C/O = 1.0 threshold the transmission spectrum is dominated by H$_2$O and HCN, and exhibits strong features of CO at 2.3 and 4.6 $\mu$m. Weak features from CH$_4$ are also seen at 3.4 and 7.6 $\mu$m.   Tight constraints on the abundances of all these molecules is therefore paramount to constrain the chemistry and C/O of these atmospheres.

Between C/O = 0.5 and 0.9 we see a gradual decrease in the molar fractions of H$_2$O and CO$_2$, and a slight increase in the  CH$_4$, HCN and C$_2$H$_2$  abundance, while CO remains relatively constant.  The resulting transmission spectra in this C/O range show the progressive decrease in the absorption of H$_2$O (which remains the dominant absorber across this C/O range) and the resulting emergence of CO, while all the other molecules remain hidden. It is only at C/O $>$ 1.0 that HCN is sufficiently abundant to be clearly seen in the transmission spectrum. We note that at this threshold we see the minimum average absorption from active gases across most of the spectrum, so that in some regions we can also  see the emergence of the collision induced absorption from H$_2$--H$_2$ and H$_2$--He pairs.

At C/O = 1.1 the H$_2$O and CO$_2$ content drastically drops, while the abundances of CH$_4$, HCN and C$_2$H$_2$ increase significantly. The corresponding transmission spectra show features of CH$_4$, HCN, CO, and C$_2$H$_2$.  At progressively higher C/O ratios we see the increase in abundance of CH$_4$, HCN, and C$_2$H$_2$, and the progressive decrease of CO abundance. However, we note that the resulting spectra are very similar to each other. The only differences in the spectra are the weakening of CO at 4.6 $\mu$m and the strengthening of C$_2$H$_2$ at 3 and 7.5 $\mu$m. 

Finally, we note that C$_2$H$_2$ might actually have additional and much stronger features than those seen here. This is because the line list used for this molecule comes from HITRAN and has been computed experimentally at Earth-like temperatures. It is therefore sub-optimal to use this line list for such high temperatures ($> 1500$ K). As an appropriate hot line list would include many more transitions resulting from the population of higher vibrational levels, additional spectral features (i.e. ``hot bands'') are expected, together with the strengthening  of the features that can already be seen at lower temperatures. Such a list is under development at ExoMol\footnote{http://www.exomol.com} (private communication).

\subsection{Retrieval of temperature profiles}

\begin{figure*}[th]
	\centering
	\includegraphics[width=510pt]{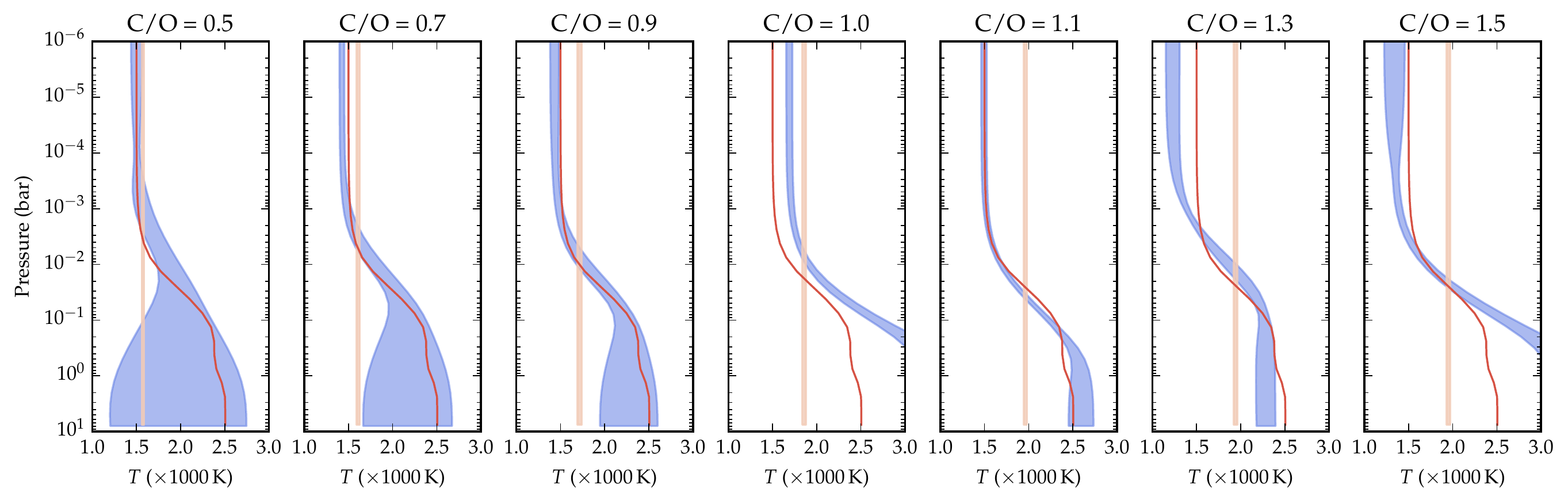}
	\caption{
Retrieved temperature profiles for the  approach with isothermal profile (pink) and parametrized profile (blue) for different C/O . The red line shows the input profile. The shaded areas show the 1 sigma confidence level.  }
	\label{fig:retrievals-tp}
\end{figure*}

Figure \ref{fig:retrievals-tp} shows the retrieved temperature profiles using the two approaches for all C/O values. It can be clearly seen that in most cases the retrieved TP profile is within 1 sigma of the input profile using the \textsc{tp-param} method, while  using the \textsc{tp-iso} method the input profile is almost entirely outside the 1 sigma retrieved error bars.

For C/O $>$ 1 (first three plots), it can be seen how the \textsc{tp-param} method fits both the upper atmosphere temperature and the lower-altitude part of the atmosphere. We found that the upper atmospheric temperature could be well fitted within about 1 to 3 sigma using the parametrized TP profile in all cases.  The high-altitude temperature was found to be $T = 1502\pm66$ K,  $T =  1425 \pm 27$ and  $T = 1433 \pm 50$ K for C/O = 0.5, 0.7 and 0.9 respectively.  Using the \textsc{tp-iso} method the the retrieved temperatures for the same C/O values were $T = 1572\pm 14$ K, $T = 1610\pm17$ K and $T = 1716\pm24$ K, respectively. In all cases, the input profile has a high altitude temperature of 1500 K.

 From these plots we can also appreciate that the non-isothermal part of the profile could be fitted within one sigma for C/O = 0.5, 0.7 and 0.9. Interestingly, we also note that for C/O $<$ 1 the constraint of the low-altitude temperature ($P > 10^{-3}$ bar) improves for higher C/O, while the fit of the high-altitude part of the profile ($P > 10^{-3}$ bar)  improves for lower C/O. 

The last three plots in Figure \ref{fig:retrievals-tp} show the retrieved temperature profiles using both approaches for C/O $>$ 1. We can see that the \textsc{tp-iso} approach retrieves a temperature of $\approx 2000$ K, with an uncertainty of $\approx 20$ K in all cases. Using the parametrized approach we could fit the high-altitude temperature within about 1 sigma for C/O = 1.1 and 1.5, and within 3.4 sigma for C/O = 1.3. We also note that while the low-altitude part of the TP profile for C/O = 1.1 and 1.3 is well constrained within about 1 sigma, for C/O = 1.5 the fit is poor for pressures higher than 0.1 bar.

For C/O = 1  we note that the TP profile is poorly retrieved, with the \textsc{tp-param} method giving slightly better results. In both cases however the input profile cannot be retrieved within several sigma: the retrieved upper atmosphere temperature is 6 and 18 sigma away from the true state using the \textsc{tp-param} and \textsc{tp-iso} methods respectively. Additionally, the lower atmosphere temperature ($P < 0.1$ bar) is not retrieved in both cases.

\subsection{Retrieval of atmospheric abundances}

\begin{figure*}
	\centering
	\includegraphics[width=400pt]{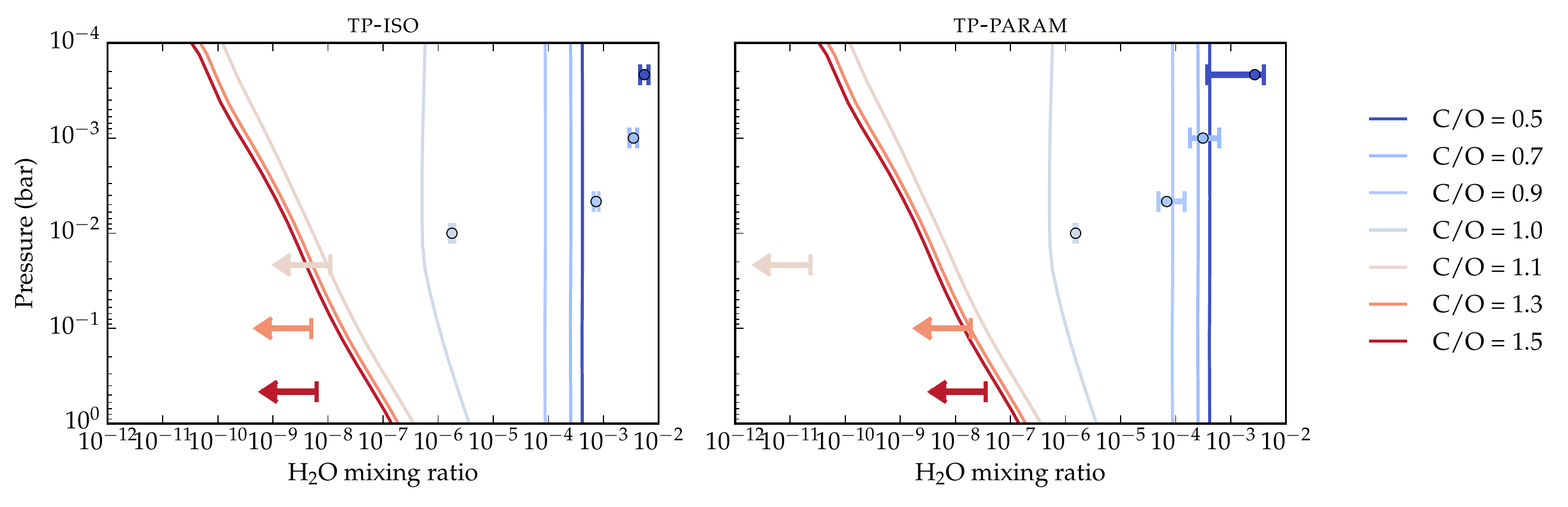}
	\includegraphics[width=400pt]{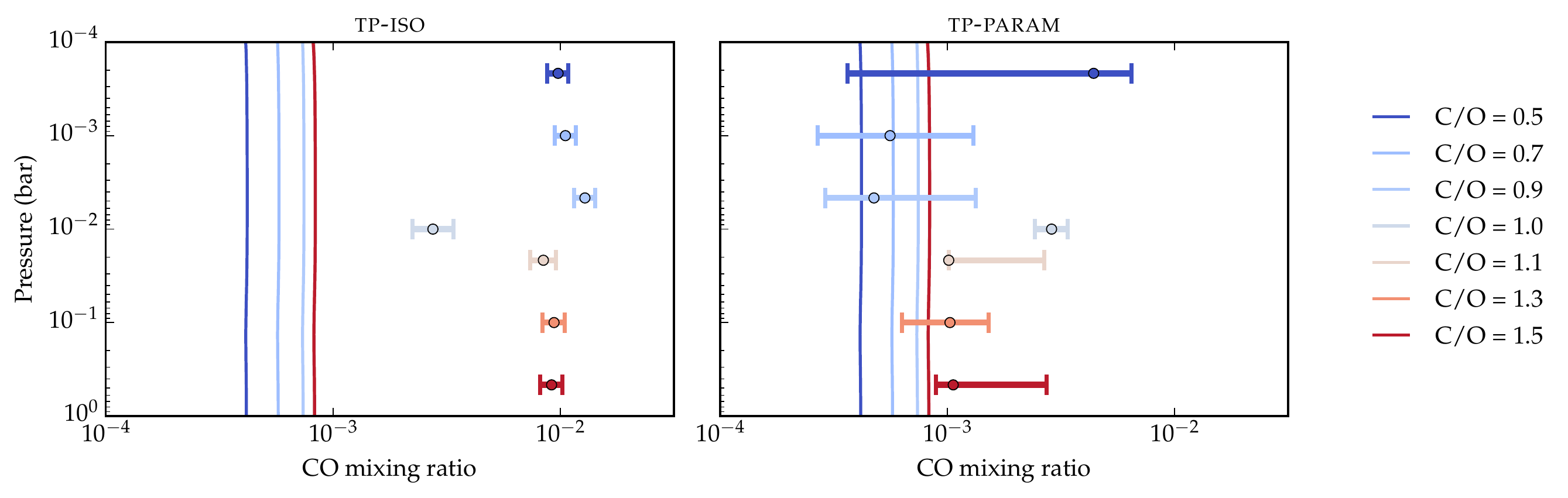}
	\includegraphics[width=400pt]{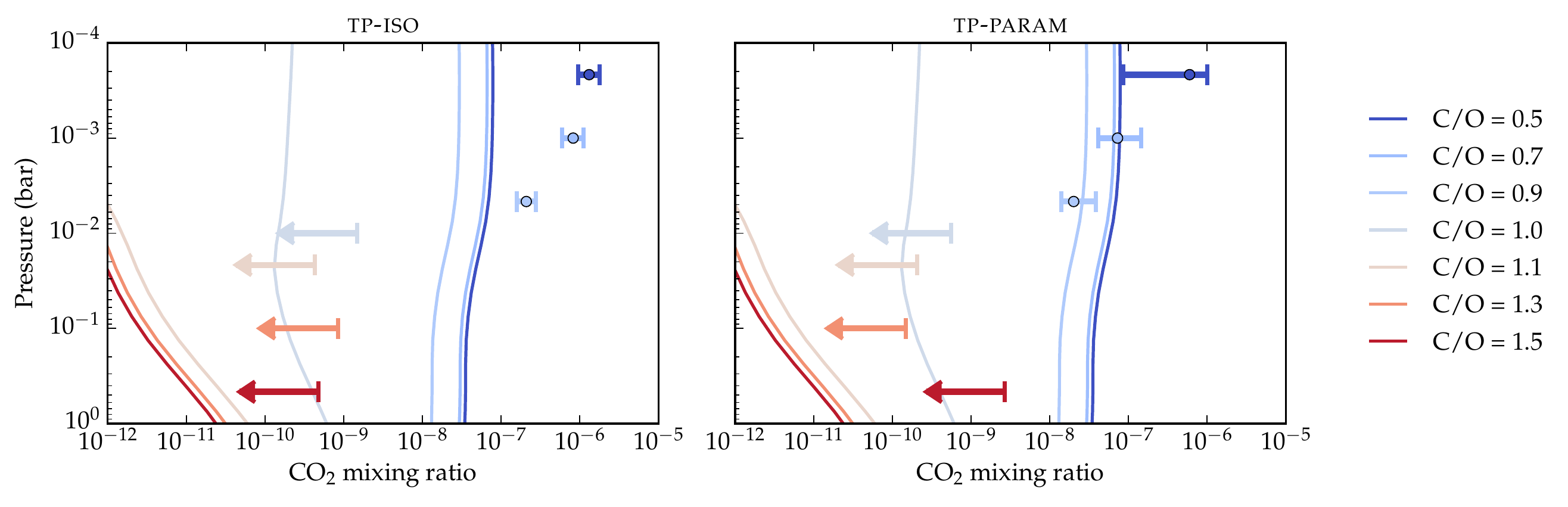}
	\caption{
Retrieved H$_2$O (top), CO (middle) and CO$_2$ (bottom) abundance for C/O = 0.5 -- 1.5 using the approach with isothermal profile (left) and parametrized TP profile (right). The solid lines show the input mixing ratio profiles for different C/O, with different colors corresponding to different C/O, as shown by the legend. The retrieved absolute mixing ratios for the different C/O are shown with error bars. Note that we retrieve constant-with-altitude mixing ratio profiles. Note also that the vertical position of the retrieved values are arbitrary.}
	\label{fig:ret1}
\end{figure*}

\begin{figure*}
	\centering
	\includegraphics[width=400pt]{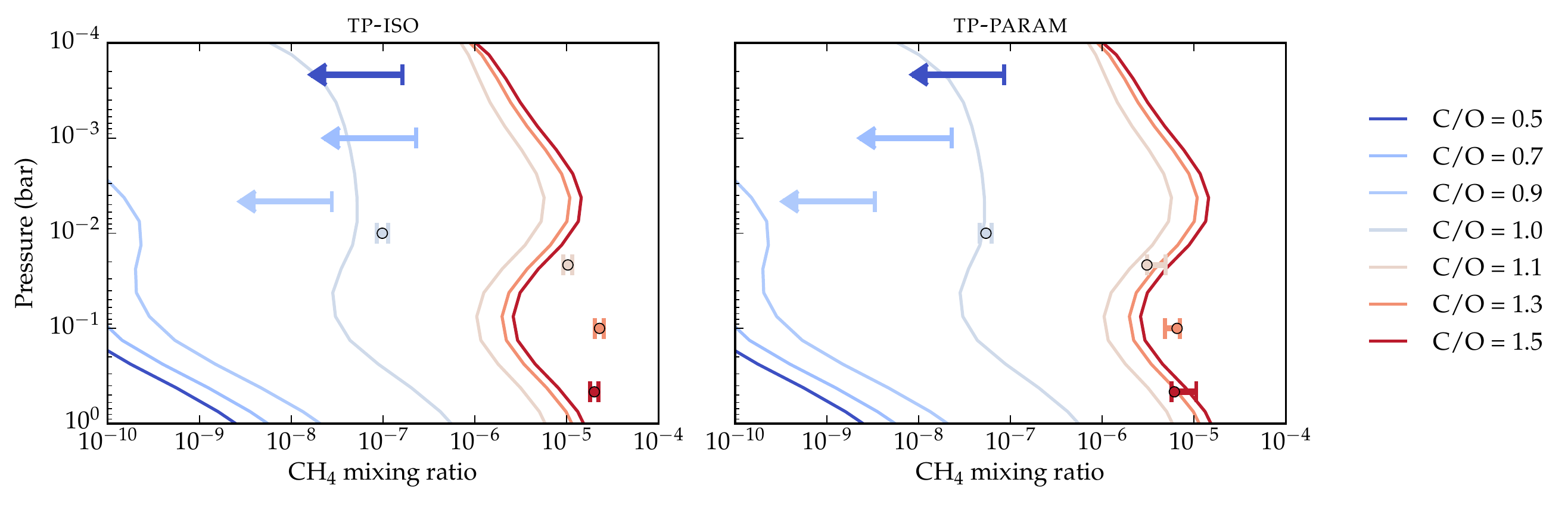}
	\includegraphics[width=400pt]{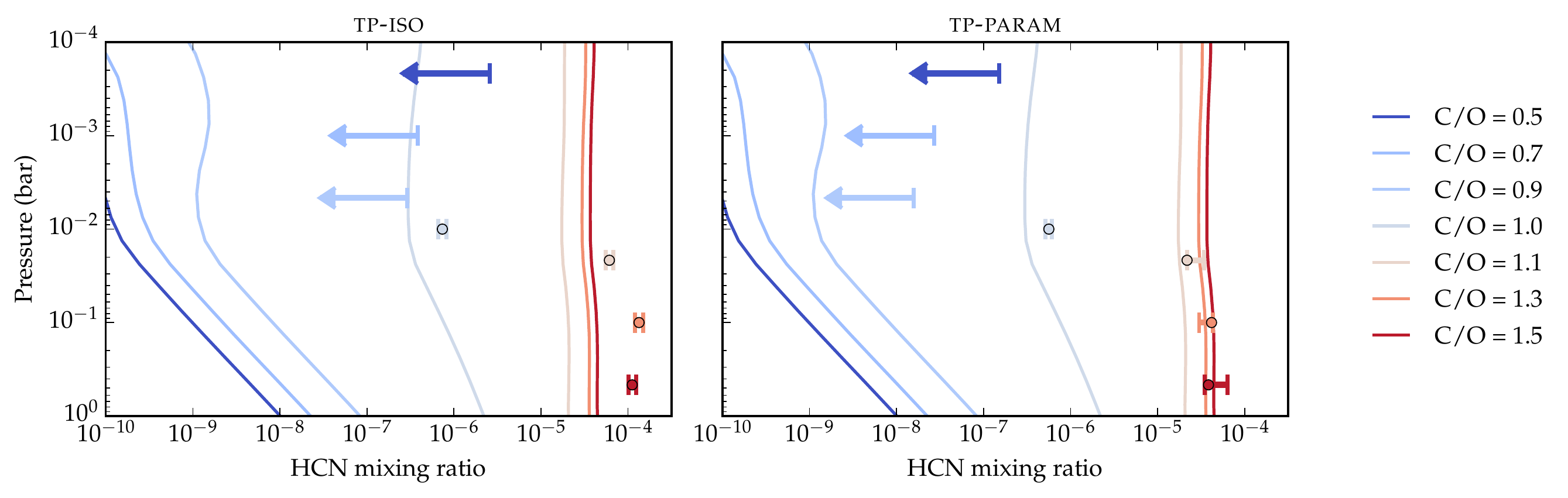}
	\includegraphics[width=400pt]{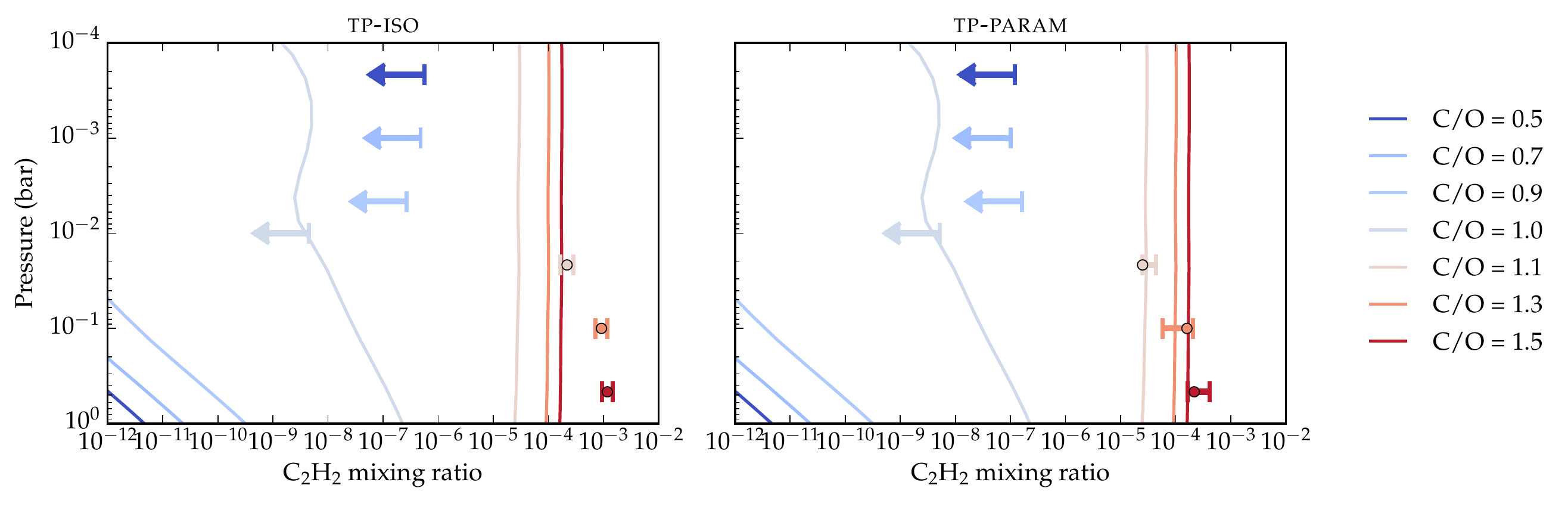}
	\caption{
	Retrieved CH$_4$ (top), HCN (middle) and C$_2$H$_2$ (bottom) abundance for C/O = 0.5 -- 1.5. Caption as in Figure \ref{fig:ret1}.}
	\label{fig:ret2}
\end{figure*}

The atmospheric retrieval results for the atmospheric abundances of H$_2$O, CO, CO$_2$, CH$_4$, HCN, C$_2$H$_2$ and NH$_3$ are shown in \autoref{tab:all_ret} in the Appendix and in Figures \ref{fig:ret1} and \ref{fig:ret2}. In these plots, the input  mixing ratios for each molecule at each C/O are also shown with solid lines as a function of pressure. We note again that the retrieved abundances are constant-with-altitude, so that a single parameter is retrieved for each molecule using both the \textsc{tp-iso} (left plots) and \textsc{tp-param} (right plots) retrieval methods. Moreover, we found that NH$_3$ is never well retrieved, hence we do not show its retrieved values in these figures. This is not surprising given that  NH$_3$ is never seen in the simulated transmission spectra (\autoref{fig:spectra}). 

\emph{In general, we found that the \textsc{tp-iso} method retrieves higher abundances by about one order of magnitude and significantly underestimates the error bars, causing strong biases, while the \textsc{tp-param} method retrieves the correct true state within one sigma for all atmospheres with C/O greater and less than 1, but not for C/O = 1. This is also because the retrieved error bars are significantly larger}.

Looking at the transmission spectra for C/O $< 1$ in Figure \ref{fig:spectra} it can be seen that  H$_2$O has  multiple features  across the entire wavelength range and is therefore the dominant molecule. Indeed, we found that, for these C/O values, the retrieved abundance of H$_2$O  has the smallest uncertainties, but only the approach using the parametrized TP profile gives unbiased results. Interestingly, the retrieval method using the isothermal approximation was found to bias the results significantly.  For example, for C/O = 0.7 the true abundance for H$_2$O at 0.1 bar is $2.5\times10^{-4}$ and is relatively constant with altitude. The retrieved abundance using the isothermal approximation was found to be $3-4\times 10^{-3}$, and 16 sigma away from the true value.  On the contrary, the retrieved abundance using the parametrized TP profile is $1.8-6.3\times10^{-4}$ and well within 1 sigma from the true value. For C/O = 0.5 and 0.9 we see similar results: using the \textsc{tp-param} method the true state is within 1 to 2 sigma of the retrieved values, but if we use the \textsc{tp-iso} method, the same retrieved values are 15 and 20 sigma away respectively from the true state. 

\begin{figure}
	\centering
	\includegraphics[width=\columnwidth]{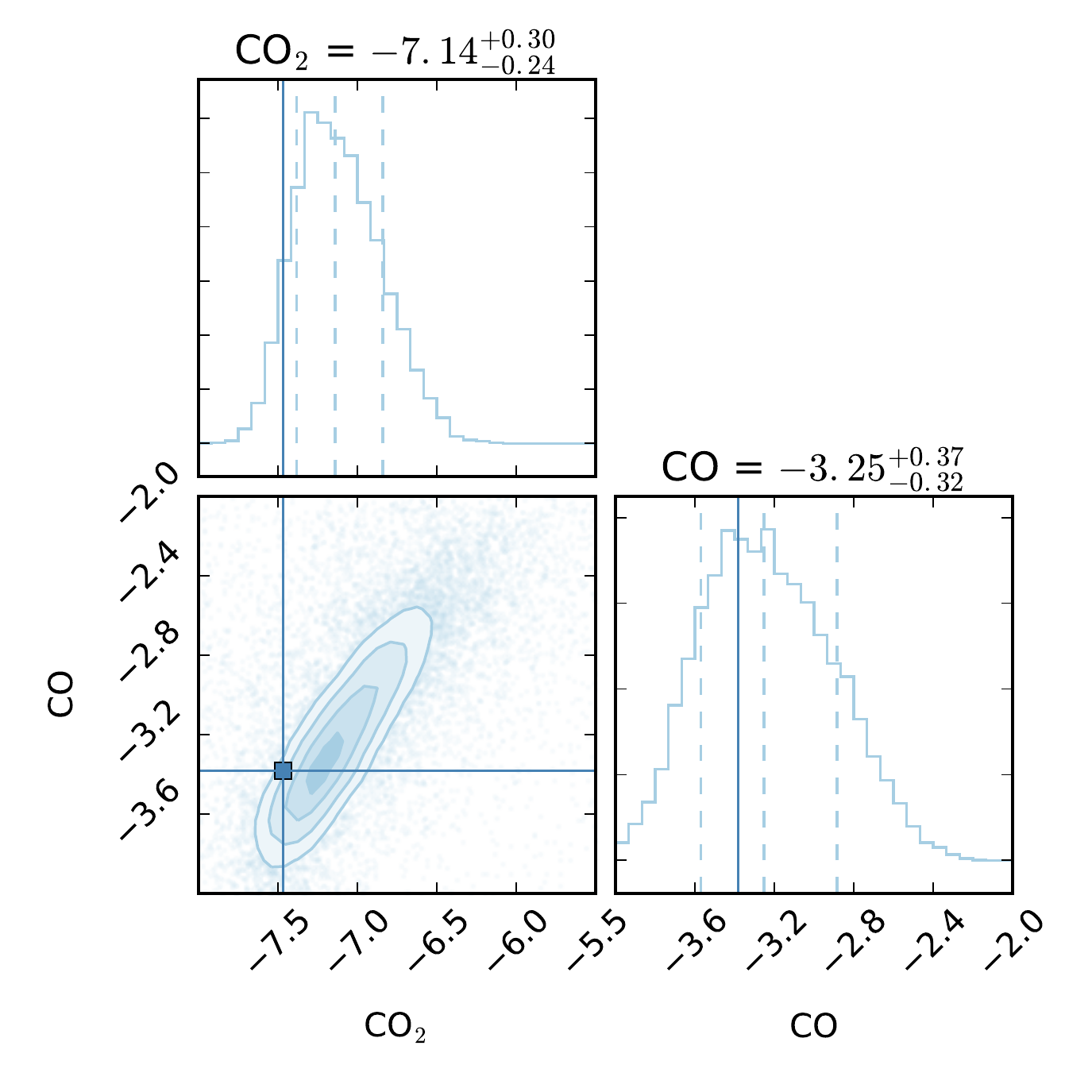}
	\caption{Posterior distributions of CO and CO$_2$ for C/O = 0.7 for the retrieval approach with a parametrized TP profile.  Dashed lines in the histogram plots show the 1 sigma confidence intervals. The true state (absolute input abundance at 0.1 bar) is shown with a blue square box and straight blue lines. Note that the mixing ratios of these two molecules is approximately constant-with-altitude in this case, as seen in \autoref{fig:chemical_models}. }
	\label{fig:co_co2_posteriors_07}
\end{figure}

The two other molecules that contribute to the spectrum, CO and CO$_2$, were found to be highly degenerate, but could be retrieved within 1 to 2 sigma using the \textsc{tp-param} method. Using the \textsc{tp-iso} method, abundances were however overestimated.  For CO$_2$ we found that using the parametrized TP profile the true state is within 2 sigma of the retrieved state for C/O = 0.5, and within 1.5 and 1.1 sigma for C/O = 0.7 and 0.9 respectively. Using the isothermal profile, we obtain retrieved values that are significantly overestimated, and are 11.0, 10.2 and 9.6 sigma away from the true state for C/O = 0.5, 0.7 and 0.9 respectively. In these hot oxygen-rich atmospheres the retrieved abundances of CO and CO$_2$ must however be interpreted with caution, as both molecules have the only detectable feature in the same wavelength range ($\approx 4.0-5.5\,\mu$m). From Figure \ref{fig:co_co2_posteriors_07}, showing the posterior distribution of CO and CO$_2$ for C/O = 0.7 using the parametrized TP-profile, it can be appreciated that the retrieved absolute abundances for these two molecules are highly degenerate. For C/O $<$ 1 no other molecules could be retrieved, and only upper limits could be obtained. 

The transmission spectra of these atmospheres with C/O $>$ 1 show that the dominant molecules are CH$_4$, HCN, C$_2$H$_2$ and CO, while all other molecules remain hidden below these stronger absorbers  (Figure \ref{fig:spectra}). Only these dominant absorbers could be retrieved, while for all other molecules only upper limits could be placed. We found that also for these carbon-rich atmospheres the \textsc{tp-param} retrieval method gives considerably better results.  

Figure \ref{fig:ret2} shows the retrieved CH$_4$, HCN and  C$_2$H$_2$ abundances. For C/O $>$ 1 we can see that the input abundance profiles change significantly as a function of pressure, especially for CH$_4$. In the case of CH$_4$ we found that the \textsc{tp-iso} method significantly overestimates the abundances. For all C/O $>$ 1 the retrieved abundances are higher than the true abundances at all pressures in the atmosphere. More reasonable results are obtained with the \textsc{tp-param} method, where the retrieved abundances are always between the maximum and minimum true abundance.   

For the same carbon-rich atmospheres, the retrieved abundances of HCN and C$_2$H$_2$  using the  \textsc{tp-param} approach are all within 1 to 2 sigma of the input abundance, while the values obtained with the  \textsc{tp-iso} are always overestimated by about one order of magnitude, and are 8 to 11 sigma away from the true state. Lastly, we note that the retrieved abundances of CO were within 1 sigma of the true state using the \textsc{tp-param} method, while using the \textsc{tp-iso} method the same values are an order of magnitude higher than the true state, and have underestimated  error bars.

This case with C/O = 1 is the most peculiar as many molecules are visible in the spectrum, and their abundance varies significantly as a function of altitude. In the case of H$_2$O, CO$_2$ and CH$_4$ the true abundance profile changes by about one order of magnitude at the typical pressures probed by transmission spectra ($10^{-1}$ -- $10^{-4}$) bar (see \autoref{fig:chemical_models}).   For H$_2$O, small differences are seen between the  \textsc{tp-param} and \textsc{tp-iso} methods. The retrieved abundances are $1.4-1.6 \times 10^{-6}$ in the first case, and $1.6-1.9 \times 10^{-6}$ in the second case, while the input profile varies between $5 \times 10^{-7}$ and $4 \times 10^{-6}$ for pressures between 1 and 10$^{-4}$ bar.  For carbon monoxide, in both cases the retrieved abundances are overestimated by about one order of magnitude, with values 6 to 7.5 sigma away from the true state. This is somewhat surprising, considering that the input profile is constant with altitude. For CO$_2$ the retrieved abundance is within 1 sigma using the  \textsc{tp-param} method, and within 2 sigma using the \textsc{tp-iso} approach. Finally, for CH$_4$ and HCN the retrieved abundances are very similar using both methods, and are found to be within the maximum and minimum abundances of the input profiles, which both vary significantly as a function of altitude. 
 
\section{Discussion}
\label{sec:discussion}

\begin{figure*}
	\centering
	\includegraphics[width=500pt]{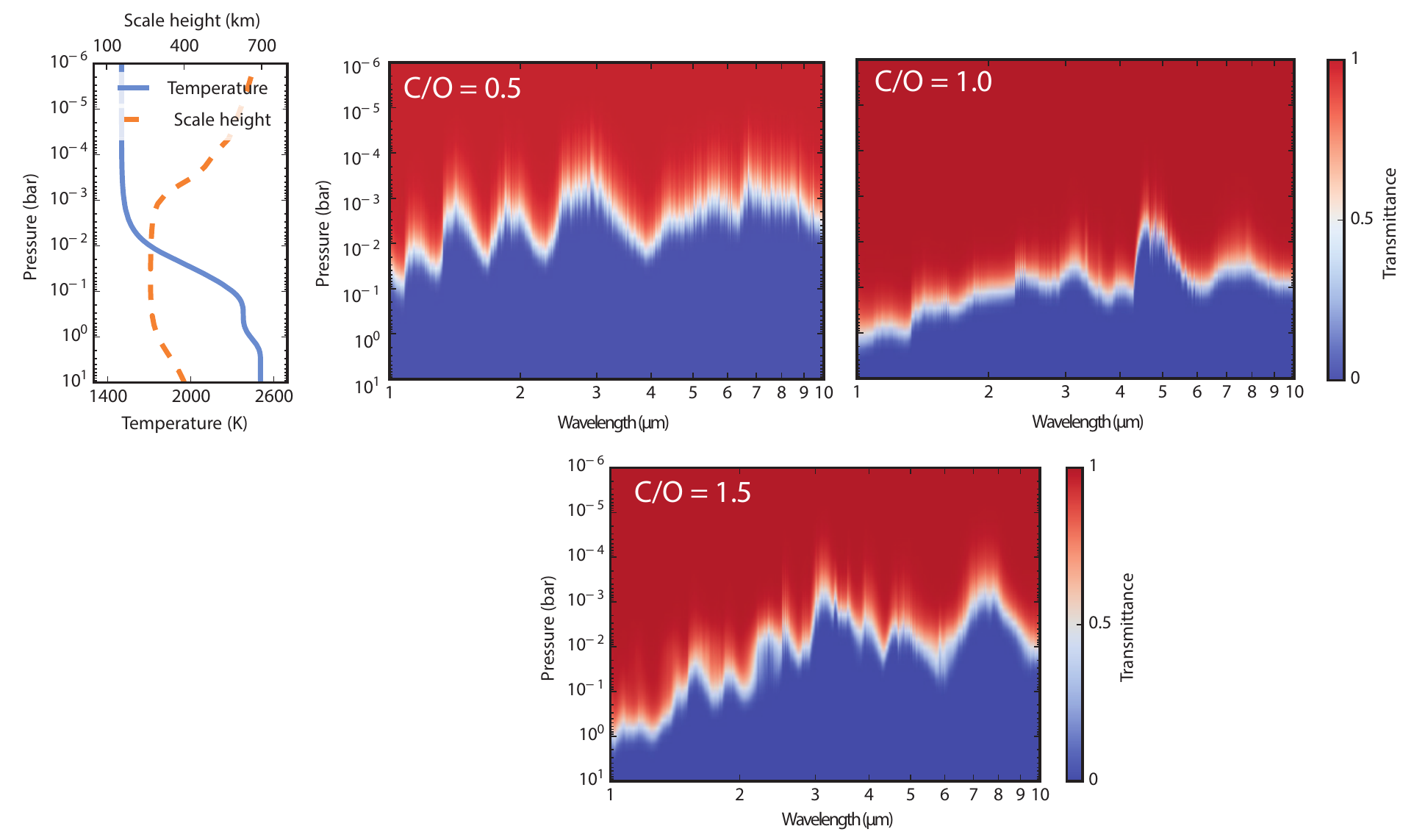}
	\caption{The first plot on the left shows the temperature profile (blue line) and scale height profile (dashed orange line) as a function of pressure. The other plots  show the spectral transmittance as a function of pressure for the models with C/O of 0.5, 1.0 and 1.5. Note that the pressure axis is the same as the first plot. The transmittance is integrated over the path parallel to the line of sight. The transmittance plots allow us to see the pressures (and therefore the temperature and scale height) probed at different wavelengths for different C/O regimes. }
	\label{fig:transmittance}
\end{figure*}

\subsection{The impact of common approximations}

The results presented in the previous section highlight how common assumptions  used in current retrieval methods for exoplanets can potentially lead to wrong conclusions.

Strong biases are seen for all C/O ratios, where we see that the isothermal approximation causes in general an overestimation of the absolute abundances by one order of magnitude, and significantly underestimates error bars.

The strongest biases are seen for H$_2$O, CO and CO$_2$ in the C/O $<$ 1 atmospheres, and for HCN, CH$_4$, C$_2$H$_2$ for the C/O $>$ 1 atmospheres. This is not surprising, given that these are the strongest absorbers for these C/O ranges, and therefore those with the smallest retrieved uncertainties.

For all these atmospheres, excluding C/O = 1, the retrieval method assuming a parametrized TP profile was found to describe the more complex temperature structure of the atmosphere, leading to retrieved values in general agreement with the true state within 1 sigma on average.   This finding opens  even new prospects for the use of this technique to characterise exoplanetary atmospheres, showing how \emph{high signal-to-noise and broad wavelength coverage transmission spectra can lead to significant constraints on the temperature profiles of the terminator region of hot Jupiter atmospheres.}

In general, the retrieval of constant-with-altitude mixing ratio profiles seems sufficient to describe the more complex real profiles when the \textsc{tp-param} approach is used, and is therefore a fair approximation in most cases. This is especially true for the C/O $<$ 1 atmospheres, where the true profiles of the most abundant molecules are constant, but it is also true for the C/O $>$ 1 atmospheres, where one of the most abundance molecules, CH$_4$, has a profile that varies significantly with altitude. The retrieved abundance of this molecule falls within the minimum and maximum true abundance, indicating that the features seen in the transmission spectra at 3.4 and 7.6 $\mu$m  probe similar pressure regions in the atmosphere. 

Retrieved parameters are more strongly affected for the C/O = 1 case, where the biases introduced by assuming a constant-with-altitude abundance profile dominate. Small differences in the retrieved values are seen using the \textsc{tp-param} and \textsc{tp-iso} methods, and the retrieved results are in both cases several sigma away from the true state. Interestingly, the TP profile retrieved using the \textsc{tp-param} method is also several sigma away from the input profile. This indicates that the biases are driven by the assumption that the abundance profiles are constant-with-altitude, which is clearly wrong for most molecules. In this case, the different features of the same molecules seen at different wavelengths (e.g. H$_2$O and CO) probe different regions of the atmosphere, where the abundances can vary significantly. Trying to fit these features using the same abundances throughout the entire atmosphere clearly leads to strong biases. We did not explore here the possibility to fit a more complex abundance profile for the molecules, but future work in this direction will be required.

The retrieved abundances obtained with the \textsc{tp-param} method will enable placing some limits on the C/O values of the observed atmospheres. Firstly, it will be clearly possible to differentiate between C/O greater or less than unity, and C/O = 1, as the spectra signatures change dramatically at this threshold. Tighter constraints on C/O can be obtained by linking the retrieved absolute abundances with atmospheric chemical models. However, our results indicate that it will be difficult. For C/O $<$ 1, the strongest tracer for C/O is water. Increasingly lower H$_2$O abundances are expected at increasing C/O, but the differences seen here are rather small, and comparable with the retrieved uncertainties (see Figure \ref{fig:ret1}). Similarly, for C/O $>$ 1, the strongest tracers are HCN and C$_2$H$_2$ (and, to a lesser extent, CH$_4$, which has however a non uniform abundance profile). However, even in this case the difference in absolute abundance is quite small, and comparable with the error bars of the retrieved values. This is not totally surprising, given that the simulated transmission spectra show very little variation between similar C/O in both the oxygen- and carbon-rich regimes. Higher signal-to-noise observations might further decrease these uncertainties, and therefore improve the inferred C/O, but we note that we are already very close to the systematic uncertainties. Other techniques might prove more effective to constrain the C/O ratio, such as emission spectra through secondary eclipse measurements and/or using chemically consistent retrieval approaches \citep[see e.g.][]{Greene:2016gx}.

\subsection{Understanding the biases}

In order to understand why, and in which scenarios, a non-isothermal profile and constant-with-altitude abundance profiles might lead to strong biases, it is instructive to look at the spectral transmittance as a function of pressure for the atmospheres under study. Figure \ref{fig:transmittance} shows the spectral transmittance integrated over the path parallel to the line of sight as a function of pressure, together with the temperature and scale height profiles. It can be seen that different spectral regions probe different pressure ranges, and therefore different temperatures and scale heights. Firstly, we note that the scale height does not increase exponentially with altitude between 10$^{-3}$ and 1 bar, as one would expect in a purely isothermal atmosphere. On the contrary, the strong temperature gradient seen at these pressures causes the scale height to stay relatively constant at $\approx$ 200 km.  For the atmosphere with C/O = 0.5  we see that most of the absorption occurs between $10^{-4}$ and $10^{-1}$ bar, while for the C/O = 1.1 case the transmission spectrum probes  higher-pressure regions, from 1 bar to $10^{-3}$ bar.  At these pressures the temperature varies from 1500 K to about 2500 K. We also note that the peak of the absorption features probe the higher-altitude and lower-temperature part of the atmospheres, while the troughs  probe the regions of the atmosphere that are almost 1000 K hotter. 

An isothermal approximation will clearly lead to several problems. Firstly, as we noted before, the scale height of an isothermal atmosphere will increase exponentially, while in this case it is roughly constant with pressure up to 1 mbar. Spectral features that probe different pressures, such as the strong water features seen for C/O $<$ 1, will therefore vary considerably if the scale height is constant with pressure or not. A second, equally important effect, is caused by the very different temperatures probed.  Molecular opacity cross sections vary considerably between the temperature regions probed here (1500 K to 2500 K), and therefore assuming a single temperature will obviously lead to further biases.

Additionally, Figure \ref{fig:transmittance} helps to explain why for the retrievals of the atmospheres with C/O $<$ 1  we found that  the fit of the low-altitude temperature improves for higher C/O, while that for the high-altitude part of the TP profile improves for lower C/O. As the C/O value increases from 0.5  to 0.9 we see that the water abundances decreases from about $4\times10^{-4}$ to $1\times10^{-4}$. The effect in the transmission spectrum is a vertical shift towards lower absorption, which also translates into a vertical shift in the transmissivity plot. This means that as the water content drops, we probe increasingly higher pressure regions of the atmospheres, meaning that we increasingly lose information from the upper-altitude part of the atmosphere. This easily explains why the uncertainty on the retrieved upper-altitude temperature of these atmospheres progressively increases, while the constraint of the temperature in the bottom layers improves for higher C/O.

So far we have only considered cloud free, broad wavelength range observations. This is the  case where common approximations are most likely to break down. Shorter wavelength ranges will for example tend to probe specific regions of the TP profiles. For instance,  an atmosphere with C/O = 1.1 observed between 1 and 3 $\mu$m will only probe pressures between 1 and 0.1 bar, where the temperature is roughly constant at $\approx 2400$ K. In this scenario, we expect the isothermal approximation to be sufficiently good. However, this is not always the case. If the atmosphere with C/O = 0.5 is observed   between 2.5 and 4 $\mu$m, we will see a strong water feature with a peak absorption coming from a region with a temperature of about 1500 K, and with  wings probing increasingly higher temperatures. Clearly, even in this case an isothermal approximation would give biased results, and our study indicates that the retrieved uncertainty of the abundance will be likely underestimated. 

The presence of uniform clouds will increase the degeneracy of model parameters, somewhat hiding the underlying biases, as the effect of a cloud deck is that of making the atmosphere opaque. A cloud deck extending to 10 mbar would for example make the atmosphere opaque to incoming radiation for pressures higher than 10 mbar. This also means that it will be impossible to probe the temperature and mixing ratio profiles in this pressure regime. In the case under study, the TP profile for pressure lower than 10 mbar is relatively isothermal, and in the presence of clouds, an isothermal approximation would therefore be appropriate.  Note however that  cloud models commonly used in current retrievals were found to cause significant degeneracies. \cite{Line:2016kv} investigated the biases of retrieving a uniform cloud cover in the presence of patchy clouds and found significant degeneracies in the retrieved mean molecular weight. 

\begin{figure*}
	\centering
	\includegraphics[width=350pt]{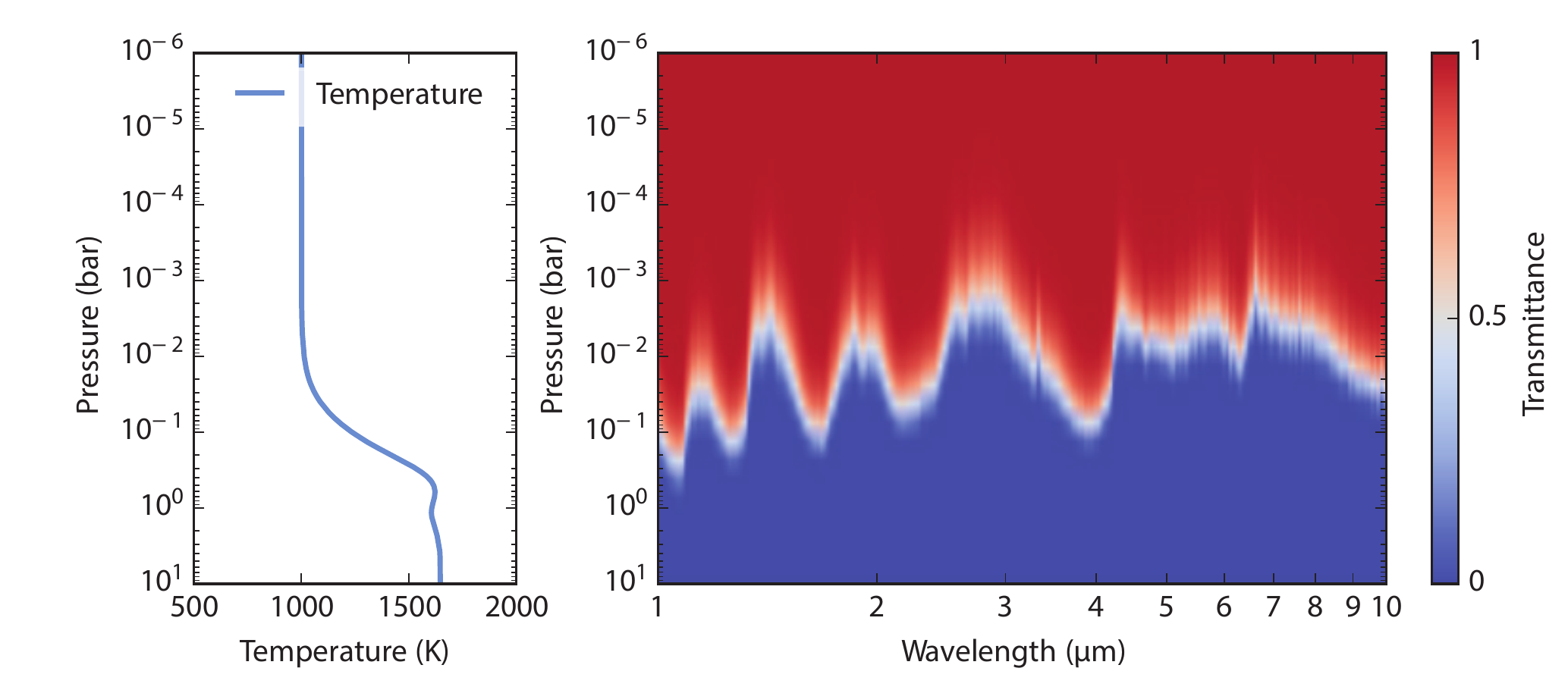}
	\caption{Temperature pressure profile and spectral transmittance for a planet with a cooler TP profile. Caption as in \autoref{fig:transmittance}. }
	\label{fig:transmittance1000}
\end{figure*}

We also note that similar biases are expected for cooler planets. \autoref{fig:transmittance1000} shows the spectral transmittance as a function of pressure for an atmosphere with a cooler TP profile, with high altitude temperature of 1000 K. The spectrum was computed from a chemical model with C/O~=~0.5, and assuming the same hot Jupiter used in this work. It can be seen that the  spectrum probes the range of pressures ($10^{-3}$ to 1~bar) where the TP profile changes more significantly. We therefore expect that the use of an isothermal profile to retrieve this spectrum will lead to similar biases to those found for the hotter planet case.

Lastly, we note that this study focused on two specific common assumptions in current retrieval methods, constant mixing ratio and temperature profiles, and the biases that these approximations can lead to. However, other strong assumptions are likely to bias our retrievals. One of the most important one is to neglect 3D dynamical effects. The simulated observations have in fact been generated using 1D chemical models, and assume a uniform chemistry and atmospheric temperature at the terminator region. Further studies that compare transmission spectra obtained with general circulation models and retrieved with the simpler 1D models are needed to address the biases of this assumption. A recent study in this direction is presented in \cite{Feng:2016uk}. 

\section{Conclusions}

In this paper we investigated the biases caused by two common assumptions in the forward models used by current retrieval methods of transmission spectra of hot Jupiter atmospheres: the use of an isothermal profile and constant-with-altitude abundances. We investigated whether these assumptions will still be valid for high signal-to-noise, broad wavelength coverage spectra such as those expected by \emph{JWST}. In order to do this, we simulated high quality observations using a chemical scheme developed by \cite{Venot:2012fr}, which include detailed temperature and abundance profiles, and we retrieved them using two simpler forward models: the first one assumes an isothermal profile (\textsc{tp-iso}), while the second one assumes a parametrized temperature profile (\textsc{tp-param}). In both cases, constant-with-altitude abundances were retrieved. We found that:

\begin{itemize}
\item The non-uniform temperature profile could be well retrieved within about 1 sigma for all cases but C/O = 1 using the \textsc{tp-param}  method. This is an important result, opening the possibility to obtain detailed temperature structure information about the terminator region of a hot Jupiter.

\item The retrieval approach that assumes an isothermal profile led to strong biases. We found that, on average, the retrieved abundances using this method are overestimated by about one order of magnitude and the error bars are underestimated. The  \textsc{tp-param} approach leads to much improved constraints, with retrieved abundances within 1--2 sigma of the input values in most cases. This is also because the retrieved uncertainties are generally larger.

\item The retrieval assumption that abundance profiles are constant-with-altitude was found to be a good approximation for C/O $<$ 1 and C/O $>$ 1 atmospheres, but not for C/O = 1. In this latter case, most of the abundance profiles have strong variations, and a uniform abundance profile is a poor approximation that leads to significant biases.  Future work will therefore be needed to address the feasibility of fitting more complex abundance profiles.

\item Although we found that differentiating between C/O $<$ 1, C/O = 1 and C/O $>$ 1 was straight forward, we also found that tighter constraints are more difficult to obtain as the differences between the transmission spectra are relatively small.  Higher signal-to-noise observations might lead to better constraints, but other biases, due to systematic uncertainties for example, might become more dominant. Emission spectra observations, possibly combined with transmission spectra, might give better constraints than transmission spectra alone.
	
\end{itemize}

These results show that when broad wavelength ranges and high signal-to-noise observations are used, the forward models used in our retrieval approaches need to allow for larger flexibility. One very simple solution is to adopt a parametrization of the temperature-pressure profile, as the one used here, but other techniques, such as the two-stage approach used in \cite{Waldmann:2015ev}, could be considered in the future.

\begin{acknowledgements}

We thank the referee for providing useful comments. This work was supported by STFC (ST/K502406/1) and the ERC projects ExoLights (617119) and ExoMol (267219). O.V. acknowledges support from the KU Leuven IDO project IDO/10/2013 and from the FWO Postdoctoral Fellowship programme. POL acknowledges support from the LabEx P2IO, the French ANR contract 05-BLAN-NT09-573739.

\end{acknowledgements}

{\small
\bibliographystyle{apj}
\bibliography{2016_Rocchetto}
}

 \newpage
\clearpage
\appendix
\begin{longtable*}{l c r r r}
	\caption{Retrieved absolute abundances with 1 sigma uncertainty for the seven molecules and seven C/O values considered in this study. For each retrieved parameter, we show in parenthesis how many sigma away the retrieved value is from the true state.} \label{tab:all_ret} \tabularnewline
		\tabularnewline
		     Parameter  & C/O & \multicolumn{1}{c}{Input value}  &  \multicolumn{2}{c}{Retrieved value} \\
		       & &   \multicolumn{1}{c}{(at 0.1 bar)}  &  \multicolumn{1}{c}{\textsc{tp-iso}}    &   \multicolumn{1}{c}{\textsc{tp-param}}   \\
		\hline
\endfirsthead
	\caption{Retrieved absolute abundances (continued).}\tabularnewline
		\hline
		     Parameter  & C/O & \multicolumn{1}{c}{Input value}  &  \multicolumn{2}{c}{Retrieved value} \\
		       & &   \multicolumn{1}{c}{(at 0.1 bar)}  &  \multicolumn{1}{c}{\textsc{tp-iso}}    &   \multicolumn{1}{c}{\textsc{tp-param}}   \\
		\hline
\endhead
H$_2$O & 0.5 & $4.08 \times 10^{-4}$ &  $4.66 \times 10^{-3}-6.53 \times 10^{-3}$ (15.4)  &  $3.73 \times 10^{-4}-1.98 \times 10^{-2}$ (1.0)  \\
 & 0.7 & $2.52 \times 10^{-4}$ &  $2.98 \times 10^{-3}-4.12 \times 10^{-3}$ (16.2)  &  $1.82 \times 10^{-4}-5.37 \times 10^{-4}$ (0.4)  \\
 & 0.9 & $8.63 \times 10^{-5}$ &  $6.61 \times 10^{-4}-8.18 \times 10^{-4}$ (20.1)  &  $4.84 \times 10^{-5}-9.75 \times 10^{-5}$ (0.3)  \\
 & 1.0 & $1.20 \times 10^{-6}$ &  $1.63 \times 10^{-6}-1.94 \times 10^{-6}$ (4.6)  &  $1.43 \times 10^{-6}-1.62 \times 10^{-6}$ (4.0)  \\
 & 1.1 & $4.56 \times 10^{-8}$ &  $<1.04 \times 10^{-8}$ (1.4)  &  $<5.44 \times 10^{-12}$ (6.2)  \\
 & 1.3 & $2.44 \times 10^{-8}$ &  $<5.13 \times 10^{-9}$ (1.5)  &  $<2.58 \times 10^{-8}$ (1.1)  \\
 & 1.5 & $1.85 \times 10^{-8}$ &  $<5.74 \times 10^{-9}$ (1.3)  &  $<1.49 \times 10^{-8}$ (0.8)  \\
CO & 0.5 & $4.13 \times 10^{-4}$ &  $8.74 \times 10^{-3}-1.10 \times 10^{-2}$ (28.0)  &  $3.64 \times 10^{-4}-5.33 \times 10^{-2}$ (0.9)  \\
 & 0.7 & $5.70 \times 10^{-4}$ &  $9.46 \times 10^{-3}-1.17 \times 10^{-2}$ (27.0)  &  $2.69 \times 10^{-4}-1.17 \times 10^{-3}$ (0.0)  \\
 & 0.9 & $7.34 \times 10^{-4}$ &  $1.15 \times 10^{-2}-1.44 \times 10^{-2}$ (25.4)  &  $2.90 \times 10^{-4}-7.79 \times 10^{-4}$ (0.4)  \\
 & 1.0 & $8.22 \times 10^{-4}$ &  $2.24 \times 10^{-3}-3.38 \times 10^{-3}$ (5.9)  &  $2.44 \times 10^{-3}-3.40 \times 10^{-3}$ (7.5)  \\
 & 1.1 & $8.24 \times 10^{-4}$ &  $7.39 \times 10^{-3}-9.62 \times 10^{-3}$ (17.6)  &  $1.01 \times 10^{-3}-1.01 \times 10^{-3}$ (799.1)  \\
 & 1.3 & $8.24 \times 10^{-4}$ &  $8.35 \times 10^{-3}-1.06 \times 10^{-2}$ (20.7)  &  $6.33 \times 10^{-4}-1.67 \times 10^{-3}$ (0.5)  \\
 & 1.5 & $8.24 \times 10^{-4}$ &  $8.17 \times 10^{-3}-1.03 \times 10^{-2}$ (20.9)  &  $8.91 \times 10^{-4}-1.26 \times 10^{-3}$ (1.4)  \\
CO$_2$ & 0.5 & $3.60 \times 10^{-8}$ &  $9.46 \times 10^{-7}-1.82 \times 10^{-6}$ (11.0)  &  $8.63 \times 10^{-8}-4.13 \times 10^{-6}$ (1.5)  \\
 & 0.7 & $3.07 \times 10^{-8}$ &  $5.95 \times 10^{-7}-1.14 \times 10^{-6}$ (10.2)  &  $4.13 \times 10^{-8}-1.27 \times 10^{-7}$ (1.5)  \\
 & 0.9 & $1.35 \times 10^{-8}$ &  $1.57 \times 10^{-7}-2.77 \times 10^{-7}$ (9.6)  &  $1.41 \times 10^{-8}-2.87 \times 10^{-8}$ (1.1)  \\
 & 1.0 & $2.10 \times 10^{-10}$ &  $<2.78 \times 10^{-9}$ (1.1)   &  $<3.24 \times 10^{-9}$ (0.3)  \\
 & 1.1 & $8.01 \times 10^{-12}$ &  $<3.32 \times 10^{-10}$ (0.6)  &  $<7.64 \times 10^{-10}$ (2.5)  \\
 & 1.3 & $4.28 \times 10^{-12}$ &  $<7.23 \times 10^{-10}$ (0.9)  &  $<8.56 \times 10^{-12}$ (0.4)  \\
 & 1.5 & $3.25 \times 10^{-12}$ &  $<4.05 \times 10^{-10}$ (1.0)  &  $<4.51 \times 10^{-7}$ (1.3)  \\
CH$_4$ & 0.5 & $6.43 \times 10^{-11}$ &  $<1.85 \times 10^{-7}$ (0.6)  &  $<1.30 \times 10^{-7}$ (0.7)  \\
 & 0.7 & $1.44 \times 10^{-10}$ &  $<2.86 \times 10^{-7}$ (0.5)  &  $<1.97 \times 10^{-8}$ (0.3)  \\
 & 0.9 & $5.39 \times 10^{-10}$ &  $<2.03 \times 10^{-8}$ (0.1)  &  $<7.50 \times 10^{-10}$ (0.5)  \\
 & 1.0 & $4.36 \times 10^{-8}$ &  $8.48 \times 10^{-8}-1.13 \times 10^{-7}$ (5.6)  &  $4.61 \times 10^{-8}-6.35 \times 10^{-8}$ (1.4)  \\
 & 1.1 & $1.16 \times 10^{-6}$ &  $9.23 \times 10^{-6}-1.15 \times 10^{-5}$ (20.0)  &  $3.06 \times 10^{-6}-3.06 \times 10^{-6}$ (43189.6)  \\
 & 1.3 & $2.20 \times 10^{-6}$ &  $2.03 \times 10^{-5}-2.55 \times 10^{-5}$ (20.4)  &  $4.85 \times 10^{-6}-8.79 \times 10^{-6}$ (3.6)  \\
 & 1.5 & $2.92 \times 10^{-6}$ &  $1.80 \times 10^{-5}-2.22 \times 10^{-5}$ (18.6)  &  $5.66 \times 10^{-6}-6.57 \times 10^{-6}$ (10.0)  \\
HCN & 0.5 & $1.32 \times 10^{-9}$ &  $<2.15 \times 10^{-5}$ (0.4)  &  $<1.15 \times 10^{-7}$ (0.0)  \\
 & 0.7 & $2.94 \times 10^{-9}$ &  $<2.94 \times 10^{-7}$ (0.1)  &  $<2.68 \times 10^{-8}$ (0.4)  \\
 & 0.9 & $1.10 \times 10^{-8}$ &  $<5.54 \times 10^{-7}$ (0.3)  &  $<3.03 \times 10^{-8}$ (0.9)  \\
 & 1.0 & $8.84 \times 10^{-7}$ &  $6.60 \times 10^{-7}-8.20 \times 10^{-7}$ (1.7)  &  $5.18 \times 10^{-7}-6.14 \times 10^{-7}$ (5.3)  \\
 & 1.1 & $2.09 \times 10^{-5}$ &  $5.53 \times 10^{-5}-6.74 \times 10^{-5}$ (10.8)  &  $2.18 \times 10^{-5}-2.18 \times 10^{-5}$ (1355.8)  \\
 & 1.3 & $3.58 \times 10^{-5}$ &  $1.20 \times 10^{-4}-1.49 \times 10^{-4}$ (12.1)  &  $3.00 \times 10^{-5}-5.79 \times 10^{-5}$ (0.5)  \\
 & 1.5 & $4.42 \times 10^{-5}$ &  $1.01 \times 10^{-4}-1.23 \times 10^{-4}$ (9.2)  &  $3.47 \times 10^{-5}-4.26 \times 10^{-5}$ (0.3)  \\
C$_2$H$_2$ & 0.5 & $8.54 \times 10^{-14}$ &  $<4.18 \times 10^{-7}$ (2.0)  &  $<9.43 \times 10^{-8}$ (2.2)  \\
 & 0.7 & $4.26 \times 10^{-13}$ &  $<5.26 \times 10^{-7}$ (1.6)  &  $<1.03 \times 10^{-7}$ (1.8)  \\
 & 0.9 & $5.96 \times 10^{-12}$ &  $<2.97 \times 10^{-7}$ (1.1)  &  $<5.24 \times 10^{-7}$ (1.3)  \\
 & 1.0 & $3.88 \times 10^{-8}$ &  $<4.08 \times 10^{-9}$ (1.6)  &  $<4.77 \times 10^{-9}$ (1.6)  \\
 & 1.1 & $2.72 \times 10^{-5}$ &  $1.66 \times 10^{-4}-2.88 \times 10^{-4}$ (7.5)  &  $2.51 \times 10^{-5}-2.51 \times 10^{-5}$ (0.1)  \\
 & 1.3 & $9.66 \times 10^{-5}$ &  $7.19 \times 10^{-4}-1.17 \times 10^{-3}$ (9.2)  &  $5.76 \times 10^{-5}-4.44 \times 10^{-4}$ (0.5)  \\
 & 1.5 & $1.68 \times 10^{-4}$ &  $9.42 \times 10^{-4}-1.47 \times 10^{-3}$ (8.7)  &  $1.64 \times 10^{-4}-2.86 \times 10^{-4}$ (0.9)  \\
NH$_3$ & 0.5 & $9.73 \times 10^{-9}$ &  $<1.04 \times 10^{-7}$ (0.5)  &  $9.74 \times 10^{-12}-3.33 \times 10^{-8}$ (0.7)  \\
 & 0.7 & $9.73 \times 10^{-9}$ &  $<4.43 \times 10^{-7}$ (0.3)  &  $<1.11 \times 10^{-8}$ (0.9)  \\
 & 0.9 & $9.65 \times 10^{-9}$ &  $<3.53 \times 10^{-6}$ (10.9)  &  $<9.63 \times 10^{-8}$ (0.5)  \\
 & 1.0 & $9.62 \times 10^{-9}$ &  $<3.76 \times 10^{-7}$ (1.4)  &  $<9.12 \times 10^{-8}$ (3.2)  \\
 & 1.1 & $8.70 \times 10^{-9}$ &  $<1.80 \times 10^{-6}$ (36.4)  &  $<2.89 \times 10^{-7}$ (4651.3)  \\
 & 1.3 & $7.97 \times 10^{-9}$ &  $<2.02 \times 10^{-6}$ (29.5)  &  $<5.93 \times 10^{-7}$ (5.6)  \\
 & 1.5 & $7.51 \times 10^{-9}$ &  $<2.15 \times 10^{-6}$ (35.7)  &  $<1.76 \times 10^{-7}$ (1452.3)  \\
		\hline
\end{longtable*}

\end{document}